# Extrinsic Morphology of Graphene

Teng Li[1]

*Department of Mechanical Engineering, Maryland NanoCenter*

*University of Maryland, College Park MD 20742*

**Abstract**

Graphene is intrinsically non-flat and corrugates randomly. Since the corrugating physics of atomically-thin graphene is strongly tied to its electronics properties, randomly corrugating morphology of graphene poses significant challenge to its application in nanoelectronic devices for which precise (digital) control is the key. Recent studies revealed that the morphology of substrate-supported graphene is regulated by the graphene-substrate interaction, thus is distinct from the random intrinsic morphology of freestanding graphene. The regulated extrinsic morphology of graphene sheds light on new pathways to fine tune the properties of graphene. To guide further research to explore these fertile opportunities, this paper reviews recent progress on modeling and experimental studies of the extrinsic morphology of graphene under a wide range of external regulation, including two dimensional and one dimensional substrate surface features and one dimensional and zero dimensional nanoscale scaffolds (e.g., nanowires and nanoparticles).

---

[1] Author to whom any correspondence should be addressed. Email: LiT@umd.edu

## 1. Introduction

The surge of interest in graphene, as epitomized by the Nobel Prize in Physics in 2010, is largely attributed to its exceptional properties [1-6]. Ultra thin, mechanically tough, electrically conductive, and transparent graphene films promise to enable a wealth of possible applications ranging from low-cost thin-film solar cells, flexible and invisible displays, to chemical and biochemical sensing arrays [7-11]. Enthusiasm for graphene-based applications aside, there are still significant challenges to their realization, largely due to the difficulty of precisely controlling the graphene properties. Graphene is intrinsically non-flat and corrugates randomly [12, 13]. Since the corrugating morphology of atomically-thin graphene is strongly tied to its electronics properties [14], these random corrugations lead to unpredictable graphene properties, which are fatal for nanoelectronic devices for which precise (digital) control is the key. Therefore, controlling the graphene morphology over large areas is crucial in enabling future graphene-based applications.

Recent studies reveal that, the *extrinsic* morphology of graphene on substrate surfaces or nanoscale scaffolds is *regulated*, distinct from the *random intrinsic* morphology of freestanding graphene. These studies on the extrinsic morphology of graphene illuminate new pathways toward fine tuning the corrugating physics, and thus the properties of graphene via external regulation. The present paper reviews recent progress on modeling and experimental studies of the extrinsic morphology of graphene, aiming to offer a knowledge base for further research to explore these fertile opportunities in controlling graphene properties. The rest of the paper is organized as follows. Section 2 discusses the random intrinsic morphology of freestanding graphene; Section 3 first reviews the regulated extrinsic morphology of graphene on natural substrate surfaces and engineered substrate surfaces with one-dimensional (1D) and two-dimensional (2D) patterned features, and then reviews the extrinsic morphology of graphene regulated by zero-dimensional (0D) and 1D nanoscale scaffolds (e.g., nanowires and nanoparticles) patterned on a substrate. Concluding remarks are given in Section 4.

## 2. Intrinsic morphology of freestanding graphene

For decades, graphene was not thought to exist until it was experimentally isolated in 2004 [1], largely due to the puzzling physical structure of graphene. On one hand, graphene is a truly 2D crystal that allows electrons to transport sub-micron distances without scattering. On the other hand, theories predicted that perfect 2D crystals could not exist, because intrinsic thermal fluctuations should destroy long-range order at any finite temperature [15-17]. The existence of 2D graphene crystals in 3D space has been attributed to their *random intrinsic corrugations*: the out-of-plane corrugations lead to increased strain energy but stabilize the random thermal fluctuation. Using transmission electron microscopy, Meyer et al observed the broadening of the diffraction peaks of suspended monolayer graphene, the distinctive evidence that graphene is non flat.[13] Further simulations showed that these random corrugations in suspended graphene are about 1 nm in amplitude and 5~10 nm in wavelength.[12, 13]

## 3.     Extrinsic morphology of supported graphene

In this section, we first review the recent experimental evidence of graphene morphology conforming to natural substrate surfaces. We next describe a general energetic framework that underpins the extrinsic morphology of graphene regulated by the substrate surface. The rest of the section includes the review of recent studies of graphene's extrinsic morphology on various engineered substrate surfaces and patterned nanoscale scaffolds.

### 3.1.     Extrinsic morphology of graphene on natural substrate surfaces

When fabricated on a substrate (e.g., $SiO_2$) via mechanical exfoliation or transfer printing, graphene also corrugates, which is often attributed to graphene's intrinsic corrugations. However, recent experiments revealed that such random corrugations could be introduced by unwanted photoresist residue under the graphene if lithographic process is used. After careful removal of the resist residue, atomic-resolution images of the graphene on $SiO_2$ showed that the graphene corrugations result from its partial conformation to the $SiO_2$ substrate (Fig. 1) [18]. High-resolution scanning tunneling microscopy further indicated that the morphology of $SiO_2$-supported graphene closely matches that of the $SiO_2$ over the entire range of length scales with nearly 99% fidelity [19]. It has been further confirmed that graphene and few-layer graphene also partially follow the surface morphology of various substrates (e.g., GaAs, InGaAs and $SiO_2$) [20-23]. These experimental evidence strongly suggest that the regulated extrinsic corrugations in substrate-supported graphene are essentially distinct from the random intrinsic corrugations in freestanding graphene. Furthermore, the substrate regulation on graphene morphology has been shown to be strong enough to prevail over the intrinsic random corrugations in graphene. For example, on an atomically-flat mica substrate, the intrinsic corrugations in graphene can be smoothed, leading to an ultra-flat extrinsic morphology of the graphene [24].

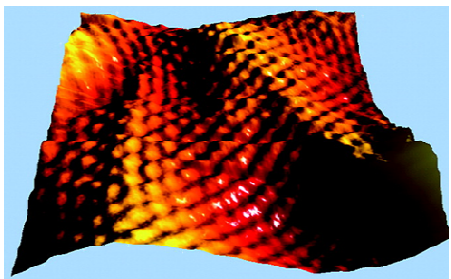

Fig. 1. Atomic resolution image of graphene partially conforming to a $SiO_2$ substrate. [18]

### 3.2. Energetics of extrinsic morphology of graphene under regulation

The extrinsic morphology of graphene regulated by the underlying substrate is governed by the interplay among three types of energies: (1) graphene strain energy, (2) graphene-substrate interaction energy, and (3) substrate strain energy.

(1) As the graphene conforms to the substrate surface morphology, the graphene strain energy increases, resulting from the out-of-plane bending as well as the in-plane stretching. Furthermore, the graphene out-of-plane deformation defines the resulting extrinsic morphology.

(2) The interaction between mechanically-exfoliated graphene and the substrate is usually weak and can be characterized by van der Waals forces. Therefore, the graphene/substrate interaction energy is given by summing all interaction energies between the graphene carbon atoms and the substrate atoms/molecules. In practice, graphene-substrate chemical bonding is also possible, which is expected to enhance the interfacial bonding. The contribution of the chemical bonding to the interaction energy is additive to that of the van de Waals bonding.

(3) The substrate strain energy depends on the substrate stiffness and the external mechanical loads. Graphene has been fabricated mostly on rigid substrates (e.g., $SiO_2$). Without external mechanical loads, the weak interaction between the ultra-thin graphene and the rather thick substrate results in negligible strain energy in the substrate. If the graphene, however, is transferred onto a flexible substrate (e.g., polymers or elastomers), and the resulting structure is subject to large deformation, the strain energy of the substrate can become comparable to that of the graphene and the graphene-substrate interaction energy, and thus needs to be considered to determine the equilibrium graphene morphology.

The energetic of graphene conforming to an underlying substrate can then be understood as follows. On one hand, as the graphene corrugates to follow the substrate surface morphology, the graphene strain energy and the substrate strain energy increases; on the other hand, by conforming to the substrate, the graphene-substrate interaction energy decreases. The total free energy of the system (denoted by the sum of the strain energy and the graphene-substrate interaction energy) minimizes, from which the equilibrium extrinsic morphology of graphene on the substrate can be determined (Fig. 2).

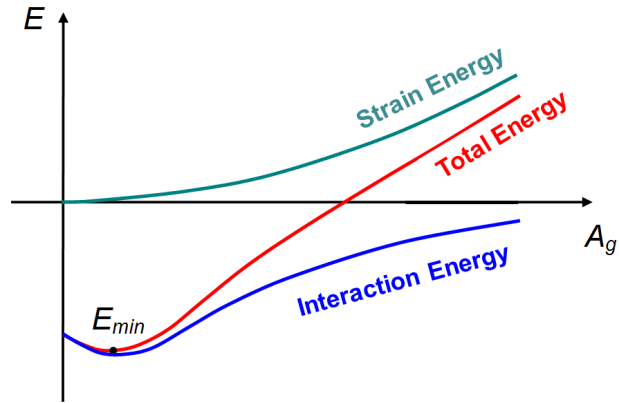

Fig. 2. Schematics of the energetics of the substrate regulation on graphene morphology. The strain energy and the graphene-substrate interaction energy are plotted as functions of the graphene corrugation amplitude $A_g$. The total free energy minimizes at an equilibrium value of $A_g$.

The above energetics sheds light on possible strategies to tailor the extrinsic morphology of graphene via external regulation. While it is difficult to directly manipulate freestanding graphene at the atomic scale, it is feasible to use mature micro/nano-fabrication techniques (e.g., nano-imprint lithography) [25-27] to pattern the substrate surface with 1D or 2D features with nano-scale precision or to form nanoscale scaffolds by patterning nanowires (1D), nanotubes (1D) or nanoparticles (0D) on a substrate surface. Graphene on such a patterned substrate surface or nanoscale scaffold will follow a regular extrinsic morphology, rather than the random intrinsic corrugations as in its freestanding counterpart. We next review recent modeling and experimental explorations of the above strategies to tailor graphene morphology, with a focus on identifying the underpinning parameters that govern the regulated extrinsic morphology of graphene, and possible morphologic features (e.g., instabilities) that could lead to new design concepts for functional components in graphene devices.

## 3.3. Extrinsic morphology of graphene regulated by patterned substrate surfaces

The energetics of graphene conforming to an underlying substrate delineated in Section 3.2 was first benchmarked by determining the extrinsic morphology of mechanically-exfoliated graphene regulated by a rigid SiO$_2$ substrate with 1D periodic surface grooves (Fig. 3a). For all simulation models hereafter in this section, we consider monolayer graphene. The substrate surface grooves have a sinusoidal profile in $x$-$z$ plane. The regulated graphene morphology is assumed to be similar to the substrate surface grooves but with a smaller amplitude. The graphene morphology and the substrate surface are described by

$$w_g(x) = A_g \cos \frac{2\pi x}{\lambda}, \quad w_s(x) = A_s \cos \frac{2\pi x}{\lambda} - h \qquad (1)$$

respectively, where $\lambda$ is the groove wavelength, $h$ is the distance between the middle planes of the graphene and the substrate surface, $A_g$ and $A_s$ are the amplitudes of the graphene morphology and the substrate surface grooves, respectively.

The graphene-substrate interaction energy is given by summing up all interaction energies due to van der Waals force between the carbon atoms in the graphene and the substrate atoms. Denoting the interaction energy potential between a graphene-substrate atomic pair of distance $r$ by $V(r)$, the interaction energy $E_{int}$ between a graphene of area $S$ and a substrate of volume $V_s$ can be given by

$$E_{int} = \int_S \int_{V_s} V(r) \rho_s dV_s \rho_C dS \qquad (2)$$

where $\rho_C$ is the homogenized carbon atom area density of graphene that is related to the equilibrium carbon-carbon bond length $l$ by $\rho_C = 4/(3\sqrt{3}l^2)$, and $\rho_s$ is the volume density of substrate atoms (i.e., the number of substrate atoms over a volume $dV_s$ is $\rho_s dV_s$) [28, 29]. Equation 2 is generally applicable to any pair potential $V(r)$. For example, Lennard-Jones (LJ) potential, $V_{LJ}(r) = 4\varepsilon(\sigma^{12}/r^{12} - \sigma^6/r^6)$, is often used to represent the graphene-substrate van der Waals force, where $\sqrt[6]{2}\sigma$ is the equilibrium distance of a graphene-substrate atomic pair and $\varepsilon$ is the bonding energy at the equilibrium distance. The interaction energy defined in Eq. (2) can be computed using a Monte Carlo numerical strategy as detailed in Ref. [30].

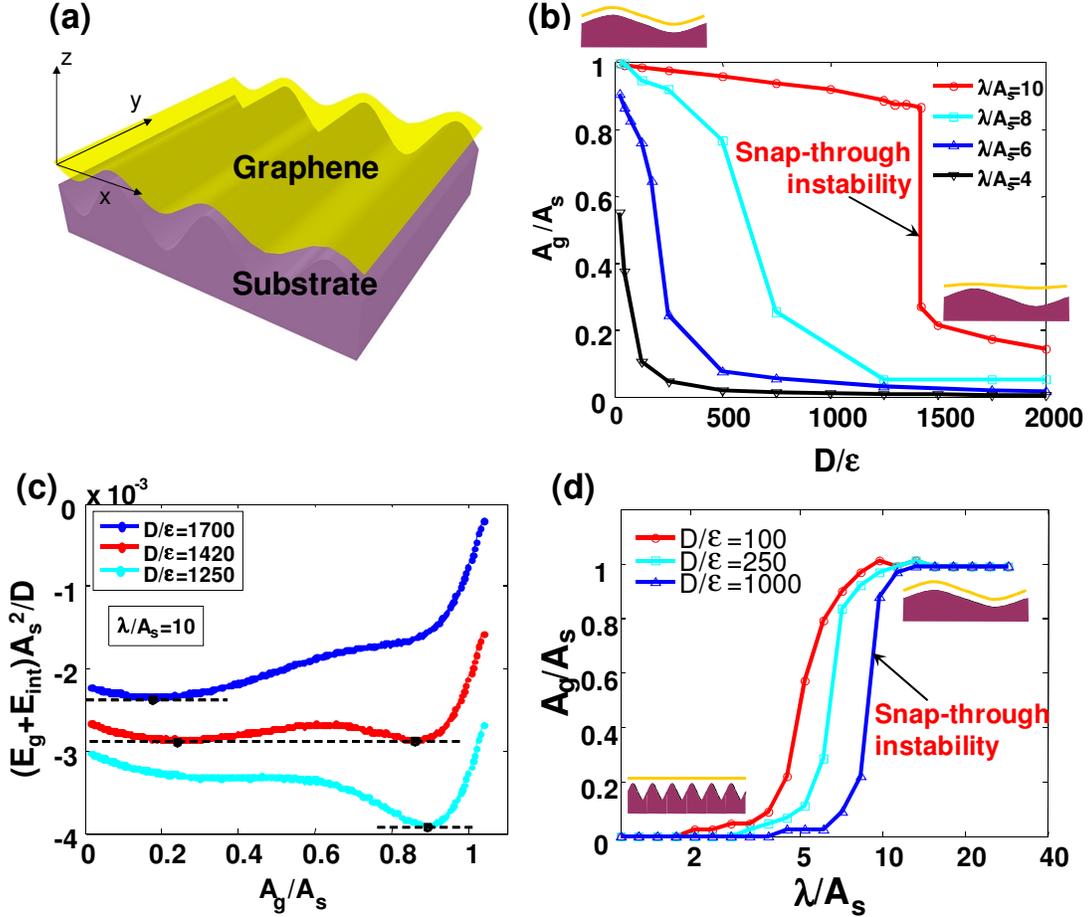

Fig. 3. (a) Schematics of a blanket graphene partially conforming to a substrate with sinusoidal surface grooves. (b) $A_g/A_s$ as a function of $D/\varepsilon$ for various $\lambda/A_s$, respectively. For $\lambda/A_s=10$ and $D/\varepsilon=1420$, the equilibrium graphene morphology snaps between two distinct states (insets): 1) closely conforming to the substrate surface and 2) nearly remaining flat on the substrate. (c) The normalized total energy as a function of $A_g/A_s$ for various $D/\varepsilon$. At a threshold value of $D/\varepsilon=1420$, $(E_g+E_{int})$ minimizes at both $A_g/A_s=0.27$ and 0.86, corresponding to the two distinct states of the graphene morphology, respectively. (d) $A_g/A_s$ as a function of $\lambda/A_s$ for various $D/\varepsilon$. The snap-through instability of the graphene morphology is also evident at a critical substrate surface roughness $\lambda/A_s$. Here $D=1.41$ eV, $\sigma=0.38$ nm. [30]

As the graphene spontaneously follows the surface morphology of the substrate (imagine a fabric conforms to a corrugated surface), the strain energy in the graphene mainly results from out-of-plane bending of the graphene, while the contribution from in-plane stretching of the graphene to the strain energy is negligible. Denoting the out-of-plane displacement of the graphene by $w_g(x, y)$, the strain energy $E_g$ of the graphene over its area $S$ can be given by

$$E_g = \int_S \frac{D}{2}\left[\left(\frac{\partial^2 w_g}{\partial x^2} + \frac{\partial^2 w_g}{\partial y^2}\right)^2 - 2(1-\nu)\left(\frac{\partial^2 w_g}{\partial x^2}\frac{\partial^2 w_g}{\partial y^2} - \left(\frac{\partial^2 w_g}{\partial x \partial y}\right)^2\right)\right]dS \tag{3}$$

where $D$ and $\nu$ are the bending rigidity and the Poisson's ratio of graphene, respectively [31-33].

By substituting $w_g(x)$ defined in Eq. (1) into Eq. (3), the graphene bending energy per unit area over a half sinusoidal period is given by

$$E_g = \frac{1}{\lambda/2}\int_0^{\lambda/2} \frac{D}{2}\left(\frac{\partial^2 w_g}{\partial x^2}\right)^2 dx = \frac{4\pi^4 D A_g^2}{\lambda^4}. \tag{4}$$

For a given substrate surface morphology (i.e., $\lambda$ and $A_s$), the graphene bending energy $E_g$ increases monotonically as $A_g$ increases (e.g., Eq. (4)). On the other hand, the graphene-substrate interaction energy $E_{int}$ is a function of $A_g$ and $h$. Due to the nature of van der Waals interaction, $E_{int}$ minimizes at finite values of $A_g$ and $h$ (e.g., Fig. 2). As a result, there exists a minimum value of $(E_g + E_{int})$ where $A_g$ and $h$ define the equilibrium morphology of the graphene on the substrate. In simulations, the equilibrium values of $A_g$ and $h$ are obtained numerically by minimizing the sum of $E_{int}$ (from Eq. (2)) and $E_g$ (from Eq. (4)).

Figures 3b show the normalized equilibrium amplitude of the graphene corrugation $A_g/A_s$ as a function of $D/\varepsilon$ for various $\lambda/A_s$, respectively. For a given substrate surface roughness (i.e., $\lambda/A_s$), if the graphene-substrate interfacial bonding energy is strong (i.e., small $D/\varepsilon$), $A_g$ tends to $A_s$. In other words, the graphene closely follows the substrate surface morphology. By contrast, if the graphene-substrate interfacial bonding is weak (i.e., large $D/\varepsilon$), $A_g$ approaches zero. That is, the graphene is nearly flat and does not conform to the substrate surface. For a given interfacial bonding energy (i.e., $D/\varepsilon$), $A_g$ increases as $\lambda/A_s$ increases. In particular, for certain range of $\lambda/A_s$ (e.g., $\lambda/A_s$ = 4 or 10), there is a sharp transition in the equilibrium amplitude of the graphene corrugation as the interfacial bonding energy varies. For example, if $\lambda/A_s = 10$, $A_g/A_s$ drops from 0.86 to 0.27, when $D/\varepsilon = 1420$. In other words, the graphene morphology snaps between two distinct states: closely conforming to the substrate surface and nearly remaining flat on the substrate surface, when the interfacial bonding energy reaches a threshold value. Such a snap-through instability of the extrinsic morphology of graphene on the substrate can be understood by the energetic understanding shown in Fig. 3c. For $\lambda/A_s = 10$, when the interfacial bonding energy is low (e.g., $D/\varepsilon = 1250$), $(E_g + E_{int})$ minimizes at $A_g/A_s$ =0.19. As $D/\varepsilon$ increases, $(E_g + E_{int})$ vs. $A_g/A_s$ curve assumes a double-well shape. At a threshold value of $D/\varepsilon$ =1420, $(E_g + E_{int})$ minimizes at both $A_g/A_s$ =0.86 and 0.27, corresponding to the two distinct states of the graphene morphology, respectively. For $D/\varepsilon$ higher than the threshold value, the minimum of $(E_g + E_{int})$ occurs at a larger $A_g/A_s$. The values of $D/\varepsilon$ in Fig. 3b represent a reasonable range of graphene-substrate interfacial bonding energy. For example, for the pair potential of C-Si, $\varepsilon$=0.00213 eV [34], which results in $D/\varepsilon$ =662.

Besides the interfacial bonding energy, the substrate surface roughness also can influence the extrinsic morphology of graphene. Figure 3d further shows the effect of substrate surface roughness $\lambda/A_s$ on the graphene amplitude $A_g/A_s$ for various values of $D/\varepsilon$. For a given interfacial bonding energy $D/\varepsilon$, there exists a threshold $\lambda_{min}$, smaller than which $A_g/A_s = 0$ (i.e., the graphene is flat, and thus not conforming to the substrate surface); and a threshold $\lambda_{max}$, greater than which $A_g/A_s = 1$ (i.e., the graphene fully conforming to the substrate surface). As $\lambda$ increases from $\lambda_{min}$ to $\lambda_{max}$, $A_g/A_s$ ramps up from zero to one. For certain range of graphene-substrate interfacial bonding energy (e.g., $D/\varepsilon > 1000$), the snap-through instability of graphene's extrinsic morphology, similar to that shown in Fig. 3b, exists, which also results from the double-well feature of the total energy profile at the threshold value of $\lambda/A_s$, similar to that shown in Fig. 3c. The dependence of $h$ on $D/\varepsilon$ and $\lambda/A_s$ (not shown in Fig. 3) can be found in Figs. 3b and 5b in Ref. [30].

In a recent study, Aitken and Huang established an analytical approach that explicitly relates the graphene-substrate interaction energy to the 1D sinusoidal surface grooves of the underlying substrate [35]. The analytical approach was also applied to predict mismatch strain induced instability of graphene morphology, that is, a compressive mismatch strain can cause a supported graphene monolayer to corrugate even on a perfectly flat substrate surface. These theoretical studies further demonstrated the tunable extrinsic morphology of graphene via substrate regulation or strain engineering. Models in Refs. [30, 35] assume that the regulated graphene morphology has the same wavelength of the substrate surface grooves. This assumption is justified if the substrate surface is modestly rough. On a severely rough substrate surface, the graphene may assume morphology of a longer wavelength to reduce the strain energy [36].

The extrinsic morphology of graphene regulated by a substrate patterned with 2D surface features (e.g., herringbone or checkerboard corrugations) has also been studied [37]. These 2D substrate surface features can be fabricated via approaches combining lithography [25, 38] and strain engineering [39, 40].

The out-of-plane herringbone corrugations of the substrate surface and the out-of-plane corrugations of the graphene regulated by such a substrate surface are described by

$$\begin{aligned} w_s &= A_s \cos\!\left((2\pi/\lambda_x)(x + A_y \cos(2\pi y/\lambda_y))\right) - h \\ w_g &= A_g \cos\!\left((2\pi/\lambda_x)(x + A_y \cos(2\pi y/\lambda_y))\right) \end{aligned}, \quad (5)$$

respectively, where $A_s$ and $A_g$ are the amplitudes of the substrate surface corrugations and the graphene corrugations, respectively; for both the graphene and the substrate, $\lambda_x$ is the wavelength of the out-of-plane corrugations, $\lambda_y$ and $A_y$ are the wavelength and the amplitude of in-plane jogs, respectively; and $h$ is the distance between the middle planes of the graphene and the substrate surface. Given the symmetry of the herringbone pattern, we only need to consider a graphene segment over an area of $\lambda_x/2$ by $\lambda_y/2$, and its interaction with the substrate. By substituting Eq. (5) into Eq. (3), the strain energy of a graphene segment over an area of $\lambda_x/2$ by $\lambda_y/2$ is given by

$$E_g = D\pi^4 A_g^2 \left(6\pi^4 A_y^4 + \lambda_y^4 + 2\pi^2 A_y^2 \left(\lambda_x^2 + 2\lambda_y^2\right)\right)/\lambda_x^3 \lambda_y^3. \quad (6)$$

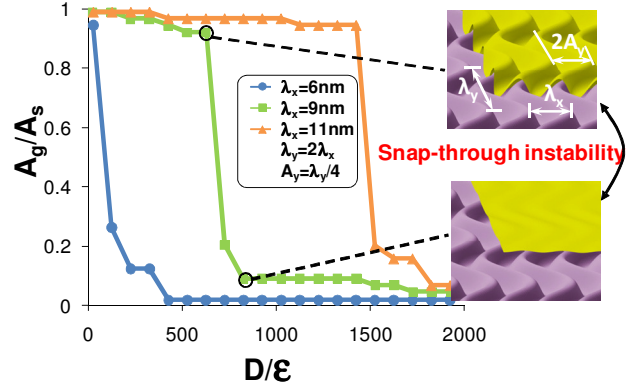

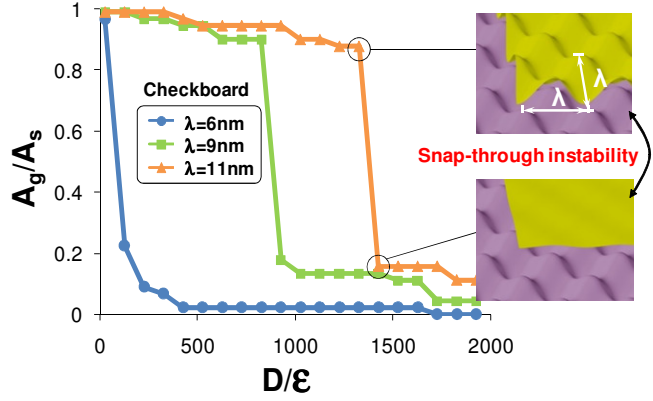

Fig. 4. (a) $A_g/A_s$ on substrates with herringbone surface corrugations as a function of $D/\varepsilon$ for various $\lambda_x$. (b) $A_g/A_s$ on substrates with checkerboard surface corrugations as a function of $D/\varepsilon$ for various $\lambda$. The insets in (a) and (b) illustrate the two distinct states of graphene morphology at the snap-through instability. [37]

As shown in Eq. (6), for a given substrate surface corrugation (i.e., $A_s$, $A_y$, $\lambda_x$ and $\lambda_y$), $E_g$ increases monotonically as $A_g$ increases. On the other hand, the graphene-substrate interaction energy, $E_{int}$, minimizes at finite values of $A_g$ and $h$, due to the nature of van der Waals interaction. As a result, there exists a minimum of $(E_g + E_{int})$ where $A_g$ and $h$ reach their equilibrium values. The computation of the graphene-substrate interaction energy and overall energy minimization can be carried out following the similar numerical strategy described above.

Figures 4a plots the normalized amplitude of the regulated graphene corrugation, $A_g/A_s$, as a function of $D/\varepsilon$ for various $\lambda_x$. Here $\lambda_y = 2\lambda_x$ and $A_y = \lambda_y/4$. Thus various $\lambda_x$ define a family of substrate surfaces with self-similar in-plane herringbone patterns and the same out-of-plane amplitude (i.e., $A_s$). For a given substrate surface pattern, if the interfacial bonding energy is strong (i.e., small $D/\varepsilon$), $A_g$ tends to $A_s$. In other words, the graphene closely follows the substrate surface (top inset in Fig. 4a). In contrast, if the interfacial bonding is weak (i.e., large $D/\varepsilon$), $A_g$ approaches zero. That is, the graphene is nearly flat and does not conform to the substrate surface (bottom inset in Fig. 4a). A snap-through instability of the extrinsic morphology of graphene, similar with that in graphene regulated by 1D substrate surface grooves, exists at a threshold value of $D/\varepsilon$, below and above which a sharp transition occurs between the above two distinct states of the graphene morphology. The threshold value of $D/\varepsilon$ increases as $\lambda_x$ increases. For a given graphene-substrate interfacial bonding energy, $A_g$ increases as $\lambda_x$ increases. That is, graphene tends to conform more to a substrate surface with smaller out-of-plane waviness. It has also been shown that, for a given interfacial bonding energy between the graphene and the substrate, similar snap-through instability of graphene exists at a critical in-plane waviness of the substrate surface. Such snap-through instability of graphene also results from the double well profile of the total free energy of the graphene-substrate system at the threshold values of interfacial bonding energy and in-plane waviness of the substrate surface.

For the case of the extrinsic morphology of graphene regulated by a substrate surface with checkerboard pattern, the substrate surface corrugations and the regulated graphene corrugations are described by

$$\begin{aligned} w_s &= A_s \cos(2\pi x/\lambda)\cos(2\pi y/\lambda) - h \\ w_g &= A_g \cos(2\pi x/\lambda)\cos(2\pi y/\lambda) \end{aligned} \quad , \quad (7)$$

respectively, where $\lambda$ is the wavelength of the out-of-plane corrugations for both the graphene and the substrate surface. The numerical strategy similar to that aforementioned was implemented to determine the regulated graphene morphology, whose dependence on the graphene-substrate interfacial bonding energy and substrate roughness has been shown to be similar to that of graphene on a substrate surface with herringbone corrugations. The snap-through of the extrinsic morphology of graphene also exists at threshold values of interfacial bonding energy and substrate roughness (e.g., Fig. 4b).

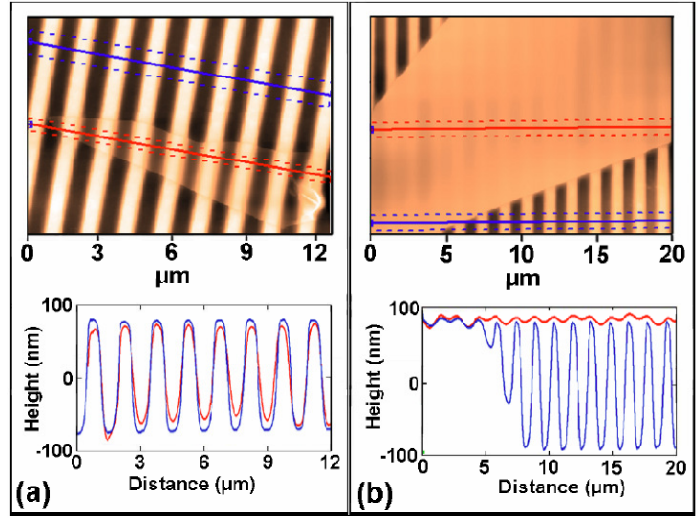

Fig. 5. Image (top) and height measurements (bottom) for (a) 8-layer and (b) 13-layer graphene on a PDMS substrate. Red lines show trajectories of scans over graphene, corresponding to red height curves (averaged between the dotted lines). Blue lines show scans of surrounding PDMS substrate. Scans over PDMS alone are taken far from graphene, to provide a baseline height unaffected by graphene. (From Ref. [41])

The extrinsic morphology of graphene regulated by a substrate with sinusoidal surface grooves as well as the snap-through instability of graphene have been recently verified in experiments [41]. In these experiments, the substrates were prepared by casting a 3 mm thick layer of polydimethylsiloxane (PDMS) onto the exposed surface of a writable compact disc. This resulted in approximately sinusoidal corrugations on the PDMS, with a wavelength of 1.5 μm and amplitude of 200nm. Graphene flakes of various thicknesses (layers) were then deposited onto the PDMS via mechanical exfoliation. The samples were first located using optical microscopy, then imaged on an atomic force microscope (AFM) which can measure the extrinsic morphology of the graphene as well as the substrate surface morphology. Figure 5 shows the images (top) and the corrugation profiles (bottom) of an 8-layer and a 13-layer graphene, respectively, on the PDMS. The 8-layer graphene can closely conform to the substrate surface grooves, while the 13-layer graphene remains nearly flat. The distinct morphology of graphene of various layers on substrate surface grooves can be readily understood by the energetic consideration delineated in Section 3.2. The strain energy of the graphene is proportional to its effective bending rigidity, which in turn approximately scales with $n^3$, where $n$ is the number of graphene layers. In other words, the strain energy increase due to an 8-layer graphene conforming to substrate surface grooves can be balanced by the decrease of graphene-substrate interaction energy; by contrast, such a strain energy increase due to a 13-layer graphene conforming to substrate surface can be too high. As a result, it stays nearly flat on the substrate. In these experiments, the compliant PDMS substrates are likely deformed near the graphene-substrate interface. However, due to the ultra-low stiffness of PDMS (about 1MPa), the resulting strain energy in PDMS is negligible when compared with that in graphene. While further experiments are desired to demonstrate the effects of substrate surface roughness and graphene-substrate interaction energy on the extrinsic morphology of graphene, the above experiments offer direct experimental evidence of substrate-regulated morphology of graphene and possible snap-through instability.

## 3.4. Extrinsic morphology of graphene regulated by 1D and 0D patterned nanoscale scaffolds

The feature length scale of the extrinsic morphology of graphene regulated by engineered substrate surfaces is limited by the resolution of nanofabrication techniques that are used to pattern the substrate surface, which is typically on the order of ten nanometers. To further explore the abundant opportunities of fine tuning the extrinsic morphology of graphene, nanoscale scaffolds with feature size approaching or comparable to the intrinsic atomic length scale of graphene become necessary to regulate the graphene morphology. The past decade has seen significant progresses in fabricating low-dimensional nanostructures (e.g., nanowires, nanotubes, and nanoparticles) with controllable size and shape [42, 43]. For example, silicon nanowires with diameter of one nanometer have been demonstrated [44]. Controllable patterning of nanowires and nanoparticles on substrate surface via self-assembly [45] or epitaxial growth [46] has been demonstrated. Nanowires or nanoparticles with diameters of down to one nanometer patterned on substrate surface offer new scaffolds to regulate graphene morphology with a resolution approaching the atomic feature size of graphene.

While the energetic described in Section 3.2 can be extended to quantitatively determine the extrinsic morphology of graphene regulated by nanowires or nanoparticles patterned on a substrate surface [47], atomistic simulations become more suitable to capture the full characteristics of the ultrafine extrinsic morphology of graphene. For example, the information of the exact positions of each carbon atom in graphene at the equilibrium extrinsic morphology, which is readily available in atomistic simulations but not in continuum modeling, becomes necessary in further first principle calculation of the resulting change in electronic properties of the graphene.

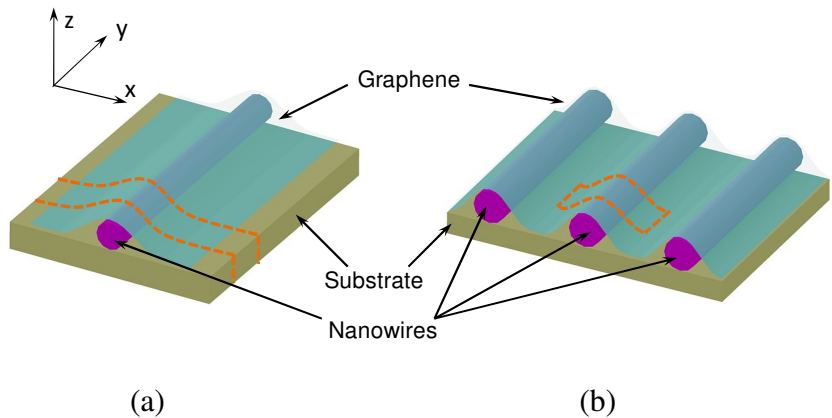

Fig. 6. Schematics of two simulation cases. (a) A graphene intercalated by a Si nanowire on a $SiO_2$ substrate; (b) A graphene intercalated by an array of Si nanowires evenly patterned in parallel on a $SiO_2$ substrate. The dashed lines delineate the portion of graphene and the underlying nanowire and substrate simulated by molecular mechanics in each case.

The extrinsic morphology of graphene regulated by a single Si nanowire on a $SiO_2$ substrate (Fig. 6a) and that by an array of Si nanowires evenly patterned in parallel on a $SiO_2$ substrate (Fig. 6a) have been studied through molecular mechanics (MM) simulations [48]. Given the periodicity of these two configurations, only the portion of the graphene marked by dashed lines and the corresponding nanowire and substrate underneath are simulated. In the MM simulations, periodic boundary conditions are applied to the two end surfaces in *y*-direction in Fig. 6a, and to the end surfaces in both *x*-and *y*-directions in Fig. 6b. The depth of the MM simulation box in *y*-direction is 30 Å and the substrate thickness is 15 Å, larger than the cut-off radius in calculating von der Waals force. The width of the graphene portion demarcated by the dash lines and that of the underlying substrate in *x*-direction, and the nanowire diameters are varied to study their effects on the graphene morphology. The C-C bonding energy in the graphene is described by the second generation Brenner potential[49]. The interaction energy between the graphene and the nanowires and that between the graphene and the substrate are computed by the sum of the van der Waals forces between all C-Si and C-O atomic pairs in the system. These two types of van der Waals forces are described by two LJ pair potentials, respectively, both of which take the general form of $V(r) = 4\varepsilon(\sigma^{12}/r^{12} - \sigma^6/r^6)$. Parameters in the C-Si pair potential and those in the C-O pair potential are listed in Table 1 in Ref. [48].

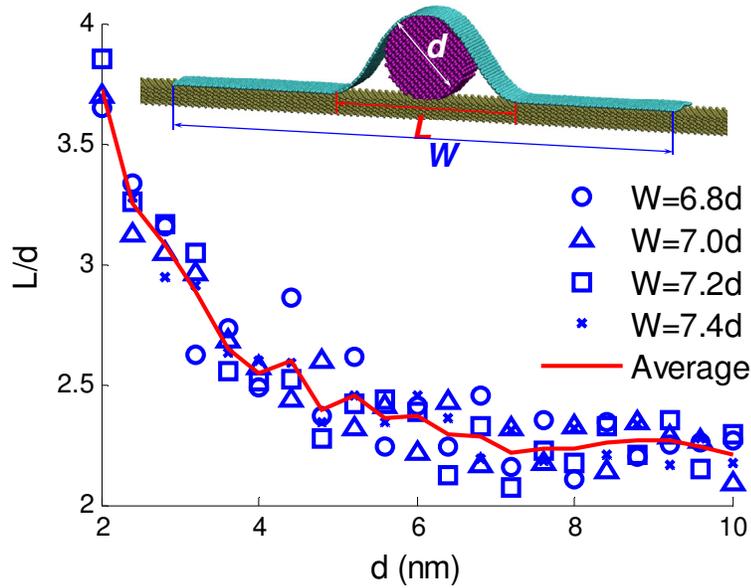

Fig. 7. MM simulation result of the extrinsic morphology of graphene intercalated by a Si nanowire on a $SiO_2$ substrate (inset). Here $d = 6$ nm and $W = 40$ nm. The data plot shows normalized width of the corrugated portion of the graphene $L/d$ as a function of $d$ for various widths of the graphene nanoribbon $W = 6.8d$, $7.0d$, $7.2d$ and $7.4d$, respectively. The solid line plots the average of the four data sets. The dash line shows the plateau value of $L/d$ when $d$ is sufficiently large. [48]

The inset of Fig. 7 shows the extrinsic morphology of graphene intercalated by a single Si nanowire on a $SiO_2$ substrate. The graphene portion far away from the Si nanowire conforms to the flat surface of the $SiO_2$ substrate while the middle portion of the graphene partially wraps around the Si nanowire. The geometry of the graphene-nanowire-substrate system at the equilibrium can be characterized by three parameters: the width of the corrugated portion of the graphene nanoribbon $L$, the width of the graphene nanoribbon $W$, and nanowire diameter $d$. Figure 7 plots $L/d$ as a function of $d$ for various widths of graphene nanoribbon $W/d = 6.8$, $7.0$, $7.2$ and $7.4$, respectively. When the graphene nanoribbon is sufficiently wide (e.g., much larger than $d$), $L$ is roughly independent of $W$, as evident with the small variation among the results for the four different values of *W*. As shown in Fig. 7, *L/d* decreases as *d* increases, and then approaches to a plateau of about 2.2 when *d* exceeds 7 nm. In other words, the morphology of the corrugated portion of the graphene nanoribbon intercalated by a sufficiently thick nanowire is approximately self-similar.

When a blanket graphene flake is intercalated by an array of Si nanowires evenly patterned in parallel on a SiO$_2$ substrate, the nanowire spacing *W* comes into play in determining the regulated morphology of the graphene flake. Emerging from the simulations are two types of morphologies of graphene at equilibrium, depending on *W* and *d*, as shown in Figure 8. If the nanowires are widely spaced (e.g., *W>>d*), the graphene tends to conform to the envelop of each individual nanowire (Figure 8a), sags down and adheres to the substrate in between neighboring nanowires. The corrugated portion of the graphene is of a width of *L* and an amplitude of $A_g$ ( ≈ *d* in this case). Figure 8b further plots *L/d* as a function of *d* for various values of *W*. For a given *W*, *L/d* increases as *d* increases in a roughly linear manner. When compared with the case of graphene nanoribbon intercalated by a single nanowire on a substrate (e.g., Fig. 7b), the width of the corrugated portion of the graphene intercalated by patterned nanowires on a substrate is much larger. This can be explained by the constraint of the portion of the graphene sagged in between neighboring nanowires and adhered to the substrate. Therefore, the graphene cannot slide easily on the substrate to conform to the envelop of each individual nanowire closely. As a result, the corrugated portion of the graphene is under modest stretch in *x*-direction.

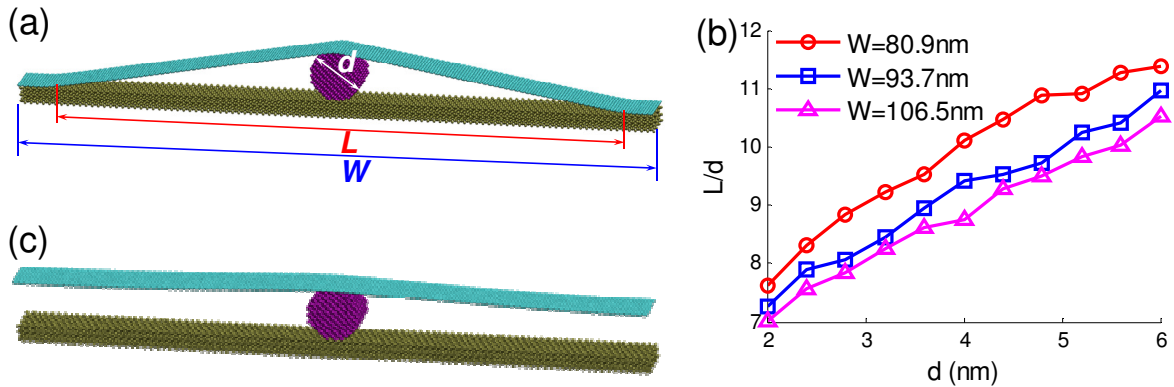

Fig. 8. MM simulation results of the extrinsic morphology of graphene intercalated by Si nanowires evenly patterned in parallel on a SiO$_2$ substrate. (a) When the Si nanowires are widely spaced (e.g., *W* is large), graphene sags in between neighboring nanowires and adhere to the substrate surface. The width of the corrugated portion of the graphene is denoted by *L*. The value of *L/d* is plotted as a function of *d* for various values of *W* in (b). (c) If the nanowire spacing is small, graphene remains nearly flat, just slightly conform to the envelop of the nanowires. Here *d* = 4 nm and *W* = 48 nm in (a) and 46 nm in (c). The sharp transition between (a) and (c) as *W* varies indicates a snap-through instability of the graphene morphology. [48]

If the spacing between the patterned nanowires is not sufficiently large, the graphene flake remains nearly flat, just slightly conforming to the envelop of the nanowires with a negligible amplitude $A_g$ (Fig. 8c), a morphology of graphene distinct from that regulated by widely distributed nanowires on a substrate (i.e., Fig. 8a). For a given nanowire diameter *d*, there is a sharp transition between these two distinct morphologies as the nanowire spacing reaches a critical value $W_{cr}$. Such a snap-through instability of the extrinsic morphology of graphene also results from the double-well energy profile of the system[30]. For the material system of graphene/Si nanowire/$SiO_2$ substrate, $W_{cr}/d$ is shown to range from 12.3 to 12.8, and is approximately independent of *d*. It has been further shown that, for a given nanowire diameter and spacing, there exists a critical graphene-substrate interaction energy, weaker than which the graphene only slightly conform to the envelop of the nanowires (e.g., $A_g/d <<1$), and stronger than which the graphene can sag in between the nanowires and adhere to the substrate (e.g., $A_g/d \approx 1$). The sharp transition between these two distinct morphologies at the critical graphene-substrate interfacial energy reveals the similar snap-through instability of the graphene morphology.

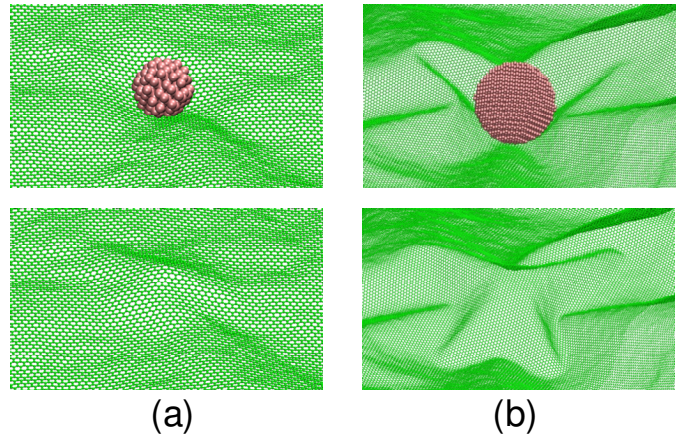

(a)          (b)

Fig. 9. MD simulation results of the extrinsic morphology of a graphene bilayer intercalated by a Si nanoparticle with diameter of (a) 2 nm and (b) 6 nm. For visual clarity, the top panel shows the nanoparticle and the bottom graphene layer, and the bottom panel shows the top graphene layer. [50]

The extrinsic morphology of a graphene bilayer intercalated by Si nanoparticles has been investigated using molecular dynamics (MD) simulation [50]. While the graphene morphology on patterned 1D Si nanowires is rather regular (e.g., forming parallel grooves), the extrinsic graphene morphology regulated by 0D Si nanoparticles presents more complicated features. For example, if the size of the Si nanoparticle is small (e.g. with a diameter of 2 nm, Fig. 9a), the graphene bilayer forms a conical dome in each layer, wrapping the nanoparticle in between. The rest portion of the graphene bilayer remains adhered by van de Waals forces between carbon atoms. As the size of the Si nanoparticle increases (e.g. with a diameter of 6 nm, Fig. 9b), both graphene layers corrugate and form ridge-like morphology. The number of ridges is same for both top and bottom graphene layer, so are the locations of the ridges. The ridge-like morphology of a graphene monolayer intercalated by silica nanoparticles (about 10 nm in diameter) dispersed on a $SiO_2$ substrate has recently been observed [51], showing similar characteristics as in the above MD simulation results. Compared to a conical dome, ridge-like morphology results in relatively smaller strain energy in the graphene layer when the intercalating nanoparticle is large. In recent experiments, it has been shown that gold nanoparticles with diameter of about 50 nm intercalating between a 2-nm-thick few-layer graphene (about 5 layers) and a $SiO_2$ substrate result in blisters in the few-layer graphene [52]. The geometry of the blisters was shown to be related to the number of gold nanoparticles wrapped underneath. For example, a single gold nanoparticle often wedges open a circular blister, while two gold nanoparticles can open up an elongated blister in the graphene. Existing experiments and modeling studies start to reveal the rich features of the extrinsic morphology of graphene regulated by 0D nanoparticles; yet more systematic investigation is desired to capture their characteristics in more details.

## 4. Concluding remarks

In this paper, we have reviewed some of the recent modeling and experimental investigations on the extrinsic morphology of graphene under a wide range of external regulation, ranging from 2D and 1D substrate surface features to 1D and 0D nanoscale scaffolds (e.g., nanowires and nanoparticles). It has been shown that the extrinsic morphology of graphene is governed by the interplay between the corrugation-induced strain energy in the graphene and the graphene-substrate (and/or graphene-nanoscafolds) interaction energy. In particular, both simulations and experiments have demonstrated the possible morphologic instability of graphene, that is, the extrinsic morphology of graphene can switch between two distinct states (i.e., closely conforming to or remaining nearly flat on the underlying substrate or nanoscaffolds) at certain critical conditions. Moreover, it has been shown that mechanical deformation (e.g., compression, torsion, bending, etc) can result in patterned morphology of carbon nanotubes [53-55]. The morphology of graphene can also be varied by mechanical deformation [56]. The quantitative understanding of the extrinsic morphology of graphene under external regulation and mechanical deformation can potentially enable controllable strain engineering of graphene, which has been shown to be able to fine tune the electronic properties of graphene [57, 58]. With the ever advancing techniques of nanofabrication and graphene synthesis over large areas, further investigations are therefore needed to fully exploit these fertile opportunities.

**Acknowledgement:**

This research is supported by National Science Foundation (grant #1069076), University of Maryland General Research Board Summer Research Award, and Maryland NanoCenter at the University of Maryland, College Park. The author acknowledges the collaboration with Z. Zhang, S. Zhu and J. Galginaitis, and is grateful for the helpful discussions with E.D. Williams, S.L. Zhang, X. Huang, N. Mason, M.S. Fuhrer, W.G. Cullen, and M. Yamamoto.